\documentclass[amsmath, amssymb, preprintnumbers, showpacs,
 showkeys,aps,pra,superscriptaddress,twocolumn]{revtex4-1}

\usepackage{color} 
\usepackage{hyperref}
\usepackage{amsmath,amssymb,mathrsfs}
\usepackage{psfrag}
\usepackage{graphicx}
\usepackage{graphics}
\usepackage{subfigure}
\usepackage{epsfig}
\usepackage{bm}
\usepackage{verbatim,color,ulem}
\usepackage{braket}
\usepackage{verbatim}

\begin{document}
	
\title{Collective modes across the soliton-droplet crossover in binary Bose mixtures}
	 
 	\author{Alberto Cappellaro}
	\email{cappellaro@pd.infn.it}
	\affiliation{Dipartimento di Fisica e Astronomia "Galileo Galilei",
		Universit\`a di Padova, via Marzolo 8, 35131 Padova, Italy}
	\author{Tommaso Macr\`i}
	\email{macri@fisica.ufrn.br}
	\affiliation{Departamento de F\'isica Te\'orica e Experimental, Universidade 
		Federal do Rio Grande do Norte, 
		and International Institute of Physics, Natal-RN, Brazil}
	\author{Luca Salasnich}
	\email{luca.salasnich@unipd.it}
	\affiliation{Dipartimento di Fisica e Astronomia "Galileo Galilei",
		Universit\`a di Padova, via Marzolo 8, 35131 Padova, Italy}
	\affiliation{CNR-INO, via Nello Carrara, 1 - 50019 Sesto Fiorentino, Italy}
	\date{\today{}}

\begin{abstract}
We study the collective modes of a binary Bose mixture 
across the soliton to droplet crossover
in a quasi one dimensional waveguide with a beyond-mean-field equation of state and
a variational Gaussian ansatz for the scalar bosonic field of the corresponding effective action.
We observe a sharp difference in the collective modes in the two regimes. 
Within the soliton regime
modes vary smoothly upon the variation of particle number or interaction strength.
On the droplet side collective modes are inhibited by 
the emission of particles. 
This mechanism turns out to 
be dominant for a wide range of particle numbers and interactions. 
In a small window of particle number
range and for intermediate interactions 
we find that monopole frequency is likely to be observed.
In the last part we focus on the spin-dipole modes 
for the case of equal intraspecies interactions and
equal equilibrium particle numbers in the presence of a weak longitudinal confinement.
We found that such modes might be unobservable in the real-time dynamics close to 
the equilibrium as their frequency is higher than 
the particle emission spectrum by at least one order of magnitude in the droplet phase.
Our results are relevant for experiments with two-component 
BECs for which we provide realistic parameters.
\end{abstract}


\maketitle

\section{Introduction}

The exquisite control of parameters in ultracold quantum gases,  
culminated with the achievement of Bose-Einstein condensation in 
alkali vapors, boosted a great experimental effort to engineer 
multi-component systems \cite{matthews1998,hall1998,hall1998-2}. 
From the theoretical standpoint, 
the spinorial nature of the order parameter and its complex dynamics, 
as in $^3 \text{He}$-$^4 \text{He}$ mixtures and three-fluids models 
\cite{khalatnikov1957,mineev1974,andreev1976}, enable an ample variety of
interesting physical effects. They span from persistent supercurrent
and internal Josephson effect \cite{beattie2013,chang2005,zibold2010},
itinerant ferromagnetism in fermionic mixtures \cite{conduit2009,salasnich2017},
to more exotic topics as the research for analogues of Hawking radiation
in bosonic mixtures \cite{liberati2006,larre2013,carusotto2017} and spin-orbit 
physics \cite{li2012,martone2012,li2013,ketterle2017}.

More recently, the theoretical proposal
of the existence of self-bound quantum droplets 
in binary bosonic mixtures by Petrov \cite{petrov2015,petrov2016} 
opened new opportunities 
for the investigation of non trivial quantum phases in BECs.
Dilute quantum droplets were first observed in a single component dipolar condensates of 
dysprosium \cite{pfau2016,pfau2016-2,ferlaino2016} 
even in absence of external confinement \cite{schmitt2016}, confirming droplet 
self stability.
Experiments with binary mixtures were recently 
performed using two internal states of $^{39}$K
in Barcelona \cite{tarruell2017-science,tarruell2017-arxiv} 
and Florence \cite{semeghini2017-arxiv} 
both in three dimensions and in quasi-one dimensional
setups.
Theoretical work on both dipolar \cite{wachtler2016,baillie2016,bisset2016,
boronat2016,cinti2016} 
and binary BECs \cite{cappellaro2017,malomed2018,nespolo2017} 
well describes such experiments in
a variety of regimes by the inclusion of leading Gaussian corrections 
to the mean field equation of state.

In a recent experiment \cite{tarruell2017-arxiv}, 
the formation of dilute 
self-bound states in a two-component BEC was
studied in a tight optical waveguide. 
Interestingly, above a critical value of the magnetic field a smooth crossover interpolating
between droplets and bright soliton states was observed.  Below such critical magnetic field and for
small particle numbers a bistable region is detected, corresponding to different minima of the energy functional.
Importantly, 
droplets appear naturally as a competition of 
meanfield and quantum fluctuation energies with opposite sign. 
Solitons, on the other hand,
are excitations appearing genuinely in low dimensional systems. 
If the 1D interaction strength is repulsive (self-defocusing nonlinearity) 
one recovers localized 
dark solitons, while for attractive condensates (self-focusing nonlinearity) one finds 
bright solitons. Both dark and bright solitons have been observed 
experimentally with atomic BECs \cite{kevrekidis07}.

In this work we study the collective modes of a two-component Bose 
gas in an optical waveguide along
the crossover from solitons to droplets.
Collective modes are an essential tool to identify and characterize 
the behavior of a quantum system
both at zero and at finite temperature.
For single component BEC with contact interactions 
in a confining potentials they have been studied 
throughly both theoretically and experimentally \cite{dalfovo1999}.
In the presence of non-local interactions, such as soft-core 
\cite{pohl2010,macri2013,macri2014,macri2014-2} 
or dipolar potentials \cite{wachtler2016-2,baillie2017}, 
or internal coupling in multi-component systems \cite{cappellaro2017,malomed2018},
 collective modes turn out to 
be crucial to detect quantum phase transitions from the uniform state 
to a structured ground state 
configuration.
Binary systems naturally support modes where the two component move in phase, 
such as monopole and 
quadrupole oscillations, which are expected to be the lowest energy excitations. 
In these simple cases the two components
share the same spatial wavefunction over the oscillation period.
More interestingly, there exits cases in which the spinorial 
nature of the order parameter allows nontrivial
collective modes where the internal components move out of 
phase around the equilibrium configuration.
The simplest case is represented by the so-called spin-dipole excitation.
For a repulsive two-component Bose gas the spin-dipole oscillation 
frequency depends crucially on the presence
of an external confining potential as well as on the 
reciprocal interaction strength and they were recently
characterized in an experiment of the Trento group \cite{ferrari2016}. 
Remarkably, for attractive binary mixtures, 
spin-dipole oscillations may take place even in the absence of an
external potential, the restoring force being proportional to the 
interspecies interaction. 
In this paper we characterize such modes highlighting 
the combined effect of the standard mean-field
attraction and the beyond-mean field corrections to the oscillation frequency. 
We employ a variational approach based on a Gaussian ansatz for the space
modes of the two components, which has been proven to perform well
close to the equilibrium configuration.
We find striking differences of the 
collective modes between droplets and solitons. 
Namely, in the soliton phase we are able to 
generalize well known results of quasi-1D BECs. 
In the droplet regime we find that collective modes are 
generically inhibited as they have energies higher than
the particle emission spectrum for a wide range of available experimental parameters.

In Section II we summarize the formalism and the variational results of the 
ground state of the two-component mixtures with equal masses.
In Section III we derive the collective modes
in the crossover from soliton to droplet for a wide range of particle numbers
and interaction strength for the experimentally 
relevant case of $^{39}$K within the variational approach.
Then, in Section IV, we focus on the spin-dipole oscillation mode, 
which is characteristic
of a two-component mixture. We specialize to the case of equal intra-species 
interaction strength for different particle numbers and compare 
it with the particle emission spectrum.
Finally, in Section V, we highlight the connection of the present study with current
experiments on mixtures and give some perspectives for future work in the Conclusions.

\section{Soliton-to-Droplet Crossover}
We begin by considering a Bose gas made of atoms in two different
hyperfine states in a volume $L^3$. Each component can be described by
a complex field $\psi_i$ ($i=1,2$), whose dynamics results from the following 
real-time low-energy effective action
\begin{equation}
S = \int dt\, d^3\mathbf{r}\Big[\sum_{j=1,2}\frac{i\hbar}{2}
\big(\psi_j^* \partial_t \psi_j -
\psi_j\partial_t \psi_j^*\big)
- \mathcal{E}_{\text{tot}}\big(\psi_1,\psi_2\big)\Big]\;.
\label{euclidean action}
\end{equation}
The total energy density $\mathcal{E}_{\text{tot}}$
reads 
\begin{equation}
\begin{aligned}
\mathcal{E}_{\text{tot}} & = \sum_{j=1,2}\bigg[\frac{\hbar^2}{2m} |\nabla\psi_j|^2+ 
V_{\text{ext}}(\mathbf{r})|\psi_j|^2 + \frac{1}{2}g_{jj}|\psi_j|^4\bigg] \\
&\qquad\qquad + g_{12}|\psi_1|^2 |\psi_2|^2 + \mathcal{E}_{\text{BMF}}
(\psi_1,\psi_2)\;,
\end{aligned}
\label{total energy density}
\end{equation}
where $V_{\text{ext}}(\mathbf{r})$ is an external confining potential and 
$g_{jk} = 4\pi\hbar^2 a_{jk}/m$, $a_{jk}$ being the intra- (with $a_{jk}\propto \delta_{j,k}$) and inter-species ($j\neq k$)
scattering lengths, $n_j=|\psi_j|^2$ is the density of the species $j$. The beyond-mean-field term $\mathcal{E}_{\text{BMF}}$ arises
from the zero-point energy of Bogoliubov collective excitations 
\cite{larsen1963,petrov2015}, namely
\begin{equation}
\mathcal{E}_{\text{BMF}} = \frac{8}{15\pi^2}\Big(\frac{m}{\hbar^2}\Big)^{3/2}
(g_{11}n_1)^{5/2} f\bigg(\frac{g_{12}^2}{g_{11}g_{22}},
\frac{g_{22} n_2}{g_{11} n_1}\bigg)
\label{lhy correction}
\end{equation}
with $f(x,y) = \sum_{\pm}[1+y\pm \sqrt{(1-y)^2 + 4xy}]^{5/2}/(4\sqrt{2})$.
In this first section, the calculation leading to the ground state properties
can be simplified by assuming the two components occupying 
the same spatial mode \cite{petrov2015,petrov2016}. The bosonic fields
can then be redefined as $\psi_j = \sqrt{n_j}\phi(\mathbf{r},t)$. 
This assumption neglects the inter-component dynamics, resulting
inadequate to probe, for example, spin-dipole oscillations. In Sec. IV, within the
variational framework, we present a modified Gaussian ansatz to include this feature
in our study. 
The minima of mean-field energy density Eq. (\ref{lhy correction})
fixes the ratio between the components of the population at which the spatial overlap 
is maximized, i.e. $N_1/N_2 = \sqrt{a_{22}/a_{11}}$ \cite{petrov2015,tarruell2017-science},
which we assume throughout this section.
By defining  $\Delta a = a_{12}+\sqrt{a_{11}a_{22}}$, 
Eq. \eqref{total energy density} can the be expressed as 
\begin{equation}
\begin{aligned}
&\mathcal{E}_{\text{tot}}  = \frac{\hbar^2 n_{\text{tot}}}{2m}
\big|\nabla\phi\big|^2
+ n_{\text{tot}}V_{\text{ext}}(\mathbf{r})\big|\phi\big|^2 \\
& + \frac{4\pi\hbar^2}{m}\dfrac{\Delta a
\sqrt{a_{22}/a_{11}}}{\Big(1+\sqrt{a_{22}/a_{11}}\Big)^2}n_{\text{tot}}
\big|\phi\big|^4 \\
& + \frac{256\sqrt{\pi}\hbar^2}{15 m} \bigg(
\frac{n_{\text{tot}}\sqrt{a_{11}a_{22}}}{1+\sqrt{a_{22}/a_{11}}}\bigg)^{5/2}
f\bigg(\frac{a_{12}^2}{a_{11}a_{22}}, \sqrt{\frac{a_{22}}{a_{11}}}\bigg)
\big|\phi\big|^5 \;,
\label{energy density with delta}
\end{aligned}
\end{equation}
where $n_{\text{tot}} = n_1 + n_2$.
Inspired by the experiment described in \cite{tarruell2017-arxiv}, 
from now on, we consider a quasi-one dimensional optical waveguide.
Then we take a harmonic confinement
on a transverse plane, so $V_{\text{ext}}= \frac{1}{2}m\omega_{\perp}^2 (x^2+y^2)$.
The presence of a harmonic potential defines a characteristic length
scale, namely $a_{\perp} = \sqrt{\hbar/(m\omega_{\perp})}$. 
In the following, all lengths are in units of $a_{\perp}$ and energies in units
of $\hbar\omega_{\perp}$. 
Scattering length will be rescaled in units of the Bohr radius $a_0$ for convenience.
The beyond-mean-field diagram and the static properties of the ground-state
configurations have been studied in \cite{tarruell2017-arxiv}.

The properties of the system can be analytically explored by
taking a Gaussian ansatz 
\begin{equation}
\phi(\mathbf{r}) = \sqrt{\frac{L^3}{\pi^{3/2}\sigma_r^2 \sigma_z}} 
\exp\bigg( - \sum_{r_i = x,y,z} \frac{r_i^2}{2\sigma_{r_i}^2}\bigg)\;
\label{gaussian ansatz}
\end{equation}
whose variational parameters are $\sigma_r$ and $\sigma_z$.
The choice of normalization factor ensures that the
original condition $||\psi||^2 = N$ is preserved. 
By replacing Eq. \eqref{gaussian ansatz} in Eq. \eqref{energy density with delta}
and taking the infinite volume limit, 
the variational energy per particle is given by \cite{tarruell2017-arxiv}
\begin{equation}
\begin{aligned}
\frac{E}{N\hbar\omega_{\perp}} & = \frac{1}{4}\Big(\frac{1}{\tilde{\sigma}_x^2}
+\frac{1}{\tilde{\sigma}_y^2}+\frac{1}{\tilde{\sigma}_z^2}\Big) 
+ \frac{\tilde{\sigma}_x^2 + \tilde{\sigma}_y^2}{4} \\
& +\frac{2 N\Delta\tilde{a}}{\sqrt{2\pi}\tilde{\sigma}_x\tilde{\sigma}_y
\tilde{\sigma}_z} 
\frac{\sqrt{\tilde{a}_{22}/\tilde{a}_{11}}}
{\Big(1+\sqrt{\tilde{a}_{22}/\tilde{a}_{11}}\Big)^2} \\
& +\frac{512\sqrt{2}}{75\sqrt{5}\pi^{7/4}} 
\frac{N^{3/2}}{(\tilde{\sigma}_x\tilde{\sigma}_x \tilde{\sigma}_z)^{3/2}} 
\bigg( \frac{\sqrt{\tilde{a}_{11}\tilde{a}_{22}}}
{1+\sqrt{\tilde{a}_{22}/\tilde{a}_{11}}}\bigg)^{5/2}
\times\\
& \qquad \times f\bigg(\frac{\tilde{a}_{12}^2}{\tilde{a}_{11}\tilde{a}_{22}}, 
\sqrt{\frac{\tilde{a}_{22}}{\tilde{a}_{11}}}\bigg)\;.
\label{variational energy}
\end{aligned}
\end{equation}
The tilde signals that we are expressing a length in units of $a_{\perp}$.
By means of Feshbach resonance \cite{errico2007}, below a critical value
of the magnetic field, the condition $\Delta a <0$ is achieved. 
\begin{figure*}[ht!]
\centering
\subfigure{\includegraphics[width=1.00\columnwidth]{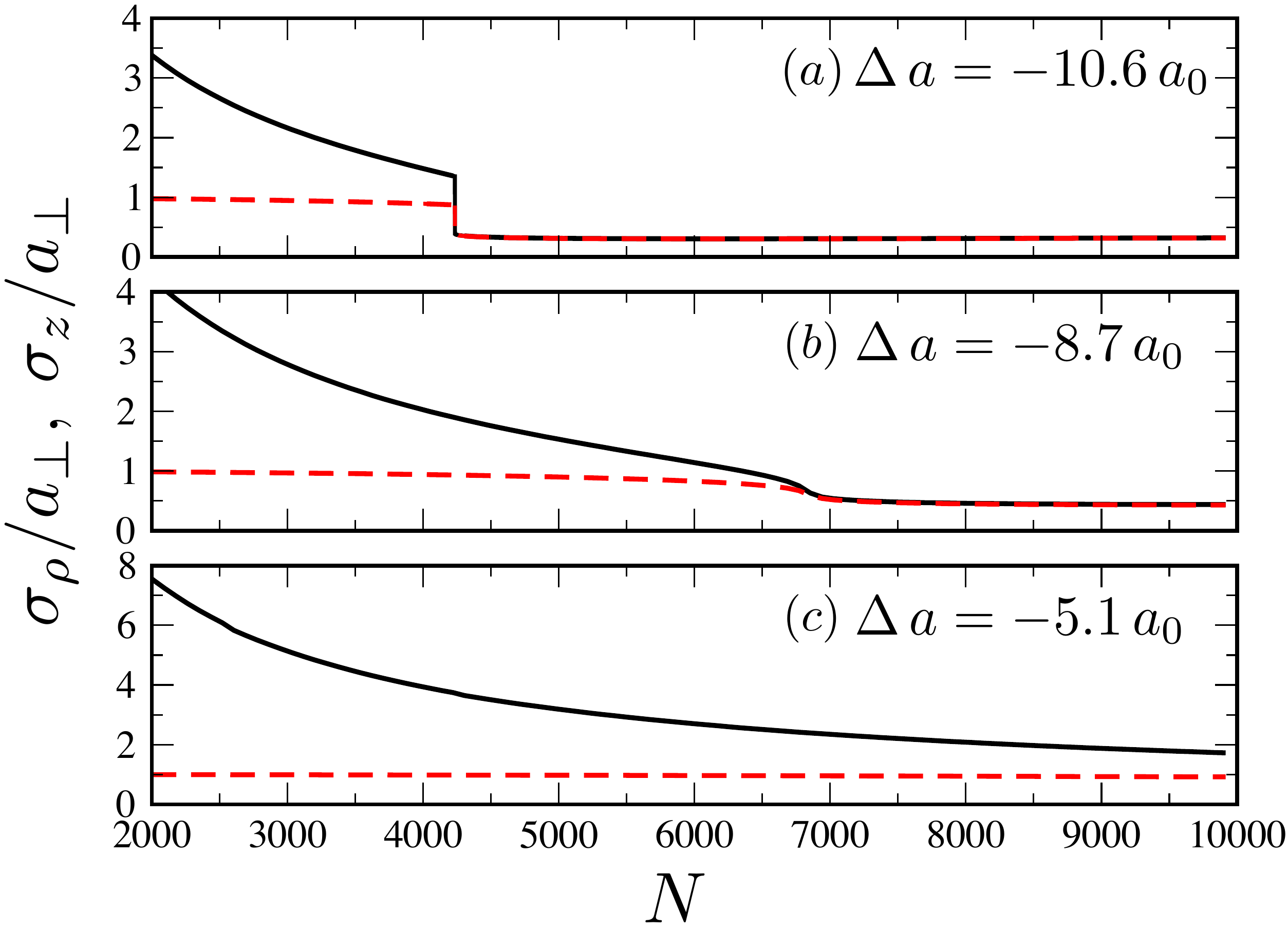}} \quad
\subfigure{\includegraphics[width=0.97\columnwidth]{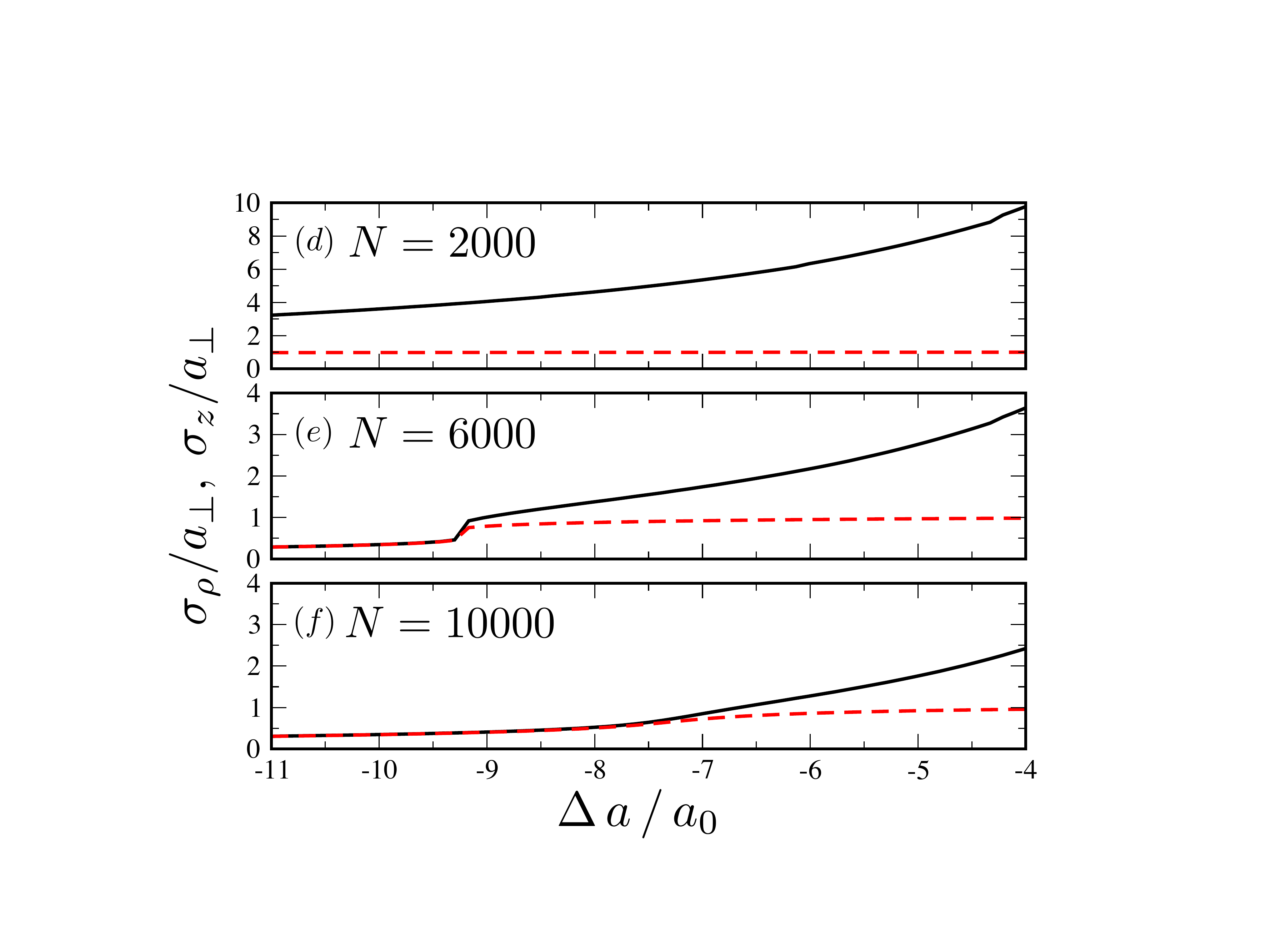}} 
\caption{Widths of the ground-state energy predicted by the variational Gaussian energy 
in Eq. \eqref{variational energy}. Widths $\sigma_i$ with $i = x,y,z$ are
expressed in units of the oscillator length $a_{\perp} =
\sqrt{\hbar/(m\omega_{\perp})}$ for several values of $\Delta a (a_0)$ (units of the Bohr's 
radius). In the plots $\sigma_z$ (solid black line), the 
gaussian width along the longitudinal direction, and $\sigma_\rho$ (red dashed line) 
along the transverse plane. Due to cylindrical symmetry of the confiment, 
$\sigma_x=\sigma_y$. 
Left panels: Variational widths as a function of the particles
number $N$ for three different values of  (a) $\Delta a=-10.6$, (b) $-8.7$, (c) $-5.1$.
In (a) a sharp change 
of $\sigma_r$ and $\sigma_z$ is observed for $N\approx 4200$
signaling a bistability. 
Right Panels:  Variational widths as a function of $\Delta a$ 
for three different values of (d) $N=2000$, (e) $6000$, (f) $10000$.
}
\label{fig1}
\end{figure*}
Here, due to the simultaneous presence of quantum fluctuations and external
confinement, two different self-bound states can be observed. First,
for low particles number or, equivalently small values of $|\Delta a|$,
the ground state of the system corresponds to the so-called solitonic state,
whose shape basically depends on the external trapping. Indeed,
$\sigma_x = \sigma_y \sim a_{\perp}$, while the longitudinal one $\sigma_z$
is much greater. It is important to remark the fact that beyond-mean-field
corrections are not necessary for the stability of this state. 
Indeed quantities such as the energy and density profile are in good agreement 
with the mean-field solutions for bright solitons in BEC 
\cite{salasnich2002}. 

The situation changes by increasing the particle number or lowering $\Delta a$.
Here, the system moves to an isotropic ground state 
($\sigma_x = \sigma_y = \sigma_z < a_{\perp}$), 
effectively independent of the confinement aspect ratios.
The existence of this self-bound state is enabled by taking into 
account the contribution of Gaussian quantum fluctuations in the variational
energy of Eq. \eqref{lhy correction}. These results are summarized
in Fig.\ref{fig1} where we plot the width as a function of N at fixed values of 
$\Delta a$ (left panels) and, in turn, as a function of $\Delta a $ for chosen values 
of N (right panels) computed via the Gaussian
ansatz Eq. \eqref{gaussian ansatz}. 

At fixed $\Delta a$, the system approaches the 
droplet state by increasing the particle number. In order to reach a
pure isotropic state, the effective mean-field attraction, i.e. $\Delta a$,
has to be strong enough. For example, at $\Delta a= -5.1\, a_0$, 
$\sigma_z$ remains two times larger than the radial width 
even at $N=10000$, while for $\Delta a = -10.6\, a_0$ the system approaches
a spherical configuration already at approximately $5000$ particles.
For weak attractive interactions (small negative $\Delta a$)
the shift from one bound state to the other occurs via a smooth crossover 
see e.g. Fig. \ref{fig1} for $\Delta a= -8.7\, a_0$.
At stronger interactions (large negative $\Delta a$) the system undergoes a bistability
where competing minima are present in the energy functional, and lead to a sharp 
structural change of the condensate as in Fig. \ref{fig1} at $\Delta a = -10.6\, a_0$ 
\cite{tarruell2017-arxiv}. A similar
picture emerges by considering a fixed particles numbers and tuning the 
effective mean-field attraction as in the right panels of Fig. \ref{fig1}.

\begin{figure*}[ht!]
\centering
\subfigure{\includegraphics[width=1.00\columnwidth]{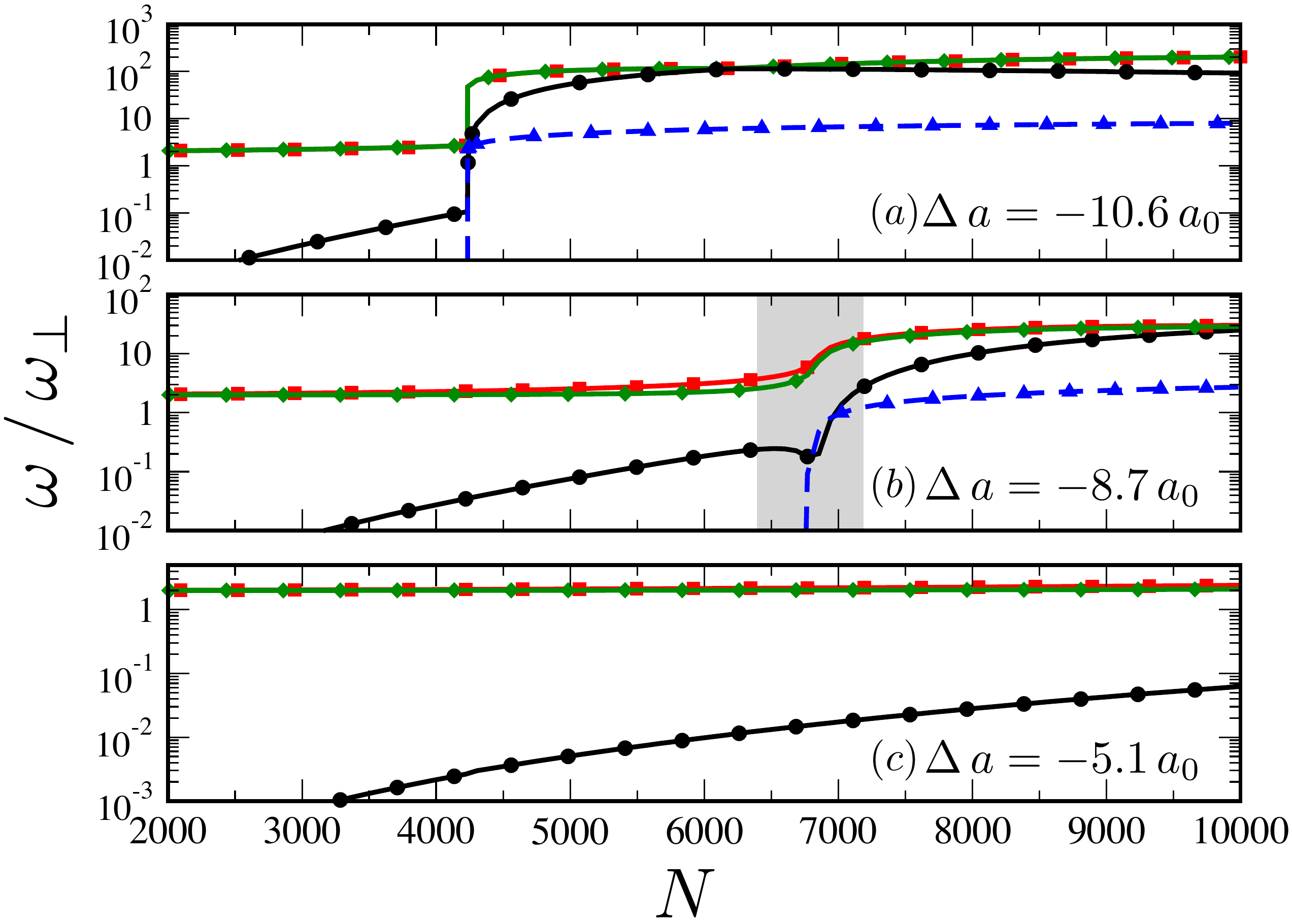}} \quad
\subfigure{\includegraphics[width=0.965\columnwidth]{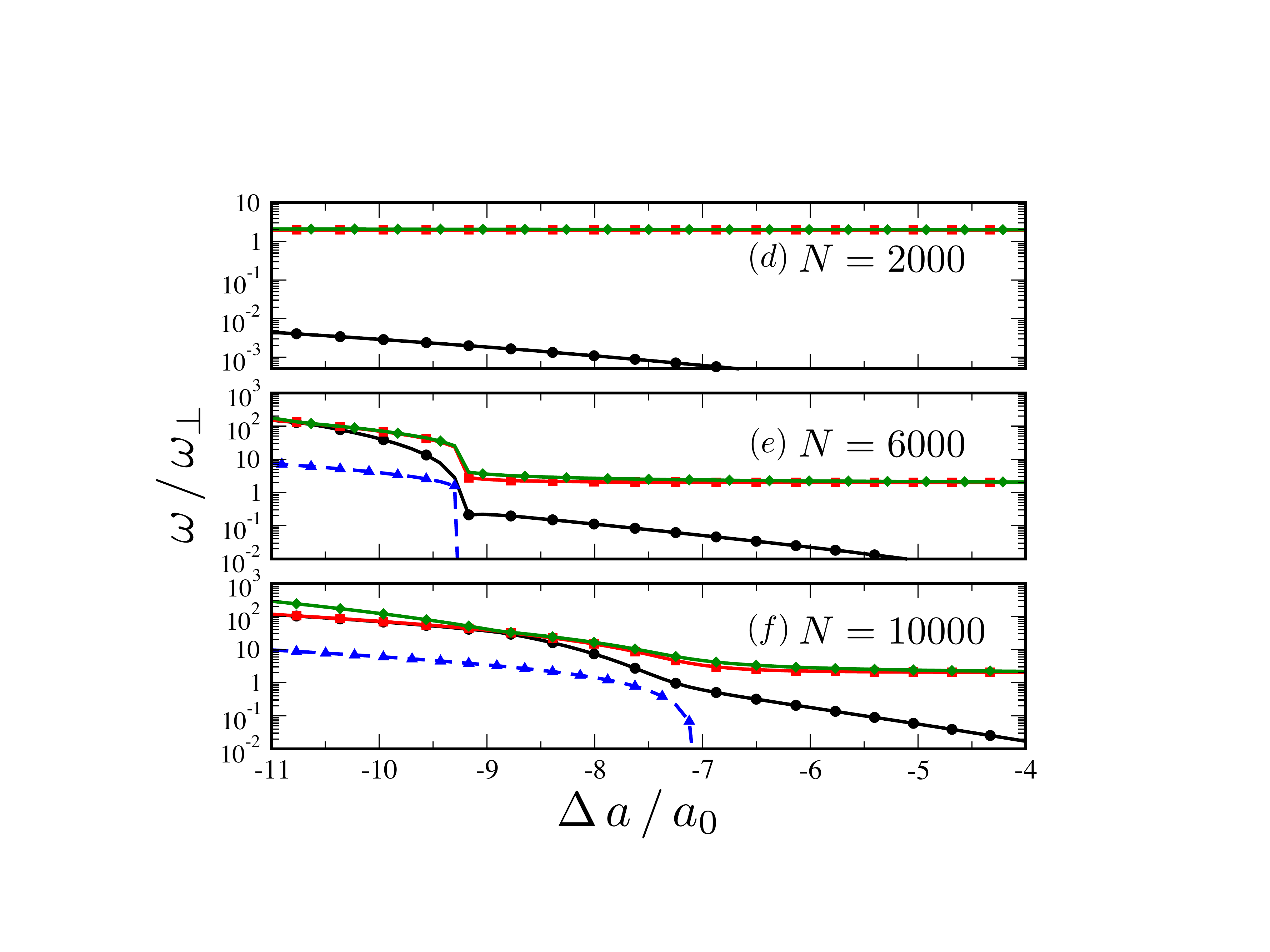}} 
\caption{Excitation frequencies $\omega_I$, $\omega_{II}$,
$\omega_{III}$ and the particle-emission threshold $-\mu$, in units of
$\hbar \omega_{\perp}$ as functions of $\Delta a$ in units of Bohr's
radius $a_0$. The emission threshold (blue triangles) is shown only in the
droplet side of the crossover. The excitation frequencies correspond to 
the eigenvalues of the Hessian matrix of Eq. \eqref{variational energy}
Left Panels: Frequencies as functions of the particle
number $N$ for three different values of the 
parameter (a) $\Delta a=-10.6$, (b) $-8.7$, (c) $-5.1$.
In (b) the grey region denotes the interval of particle numbers where the monopole mode
is lower than the particle emission spectrum in the droplet phase. 
See also Fig.\ref{fig3} for a
magnification of this area.
Right Panels: Frequencies as functions of $\Delta a$ 
for three different values of $N=2000$, (e) $6000$, (f) $10000$.
}
\label{fig2}
\end{figure*}

\begin{figure}[ht!]
\centering
\includegraphics[width=1.0\columnwidth]{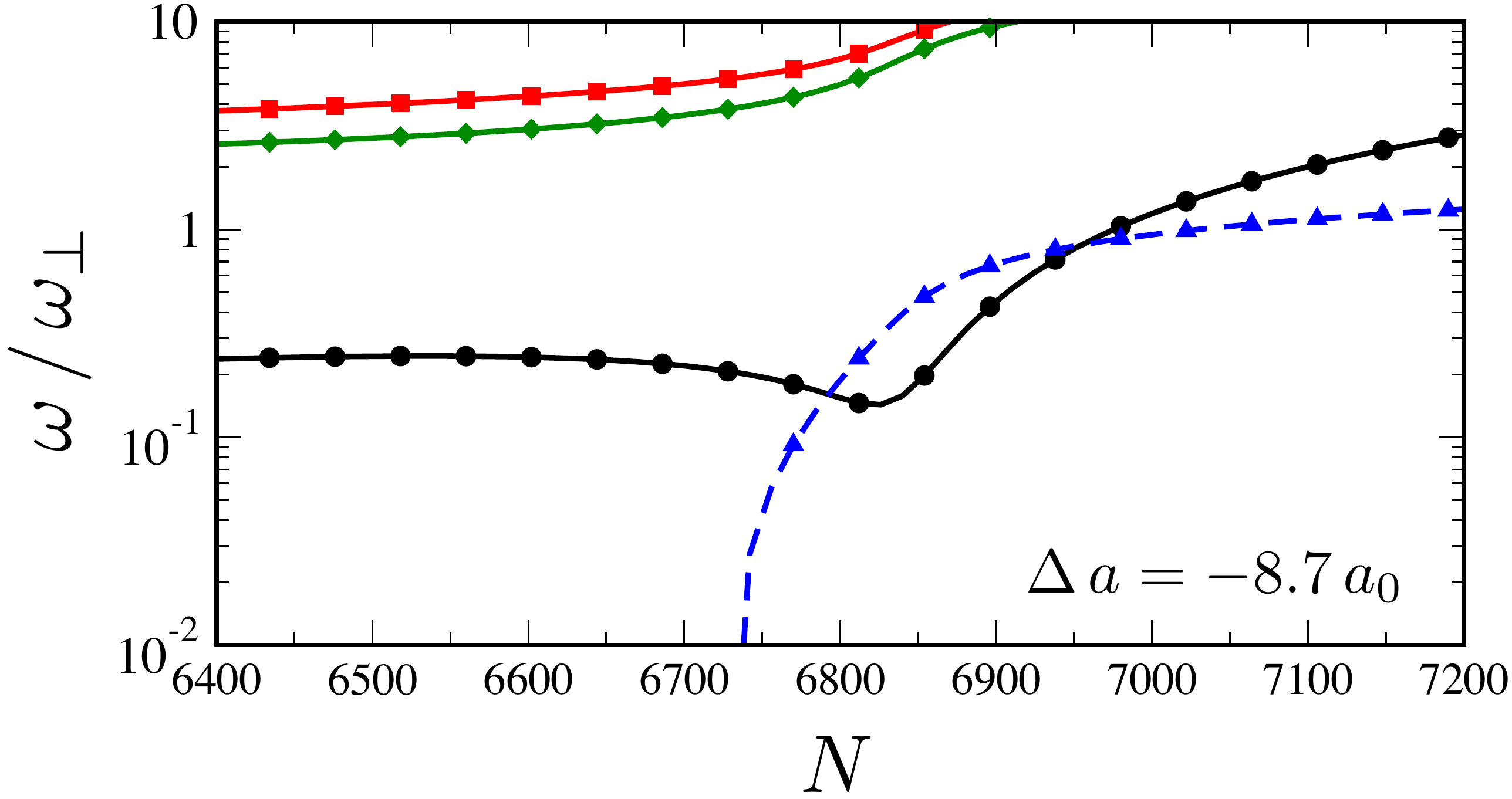}
\caption{\textit{Excitation spectrum at the crossover 
soliton-droplet at $\Delta a=-8.7\,a_{ho}$.}
Magnification of the grey region of Fig.\ref{fig2}(b) with $N=6400-7200$.
In the range $6780\lesssim N \lesssim 6950$ monopole oscillations (black line-dots) 
are higher than the particle emission spectrum (dashed blue line - triangles).}
\label{fig3}
\end{figure}

\section{Collective modes in the variational approach}
A deeper insight into the differences between solitonic and droplet states
can be reached by examining collective excitations frequencies 
around the minima of the variational energy of Eq. \eqref{variational energy},
given the Gaussian ansatz in Eq. \eqref{gaussian ansatz}.
The variational Gaussian approach results to be particularly reliable in 
predicting collective excitations frequencies for a wide range of 
superfluid systems. As an example, in \cite{malomed2018} a detailed analysis 
has been carried out by comparing the variational approach with 
numerical solutions of the Gross-Pitaevskii in purely 1D binary Bose mixtures.
Concerning excitation frequencies,
numerical and analytical predictions are shown to be in great agreement.
Moreover, our theoretical framework was also applied to study 
droplet-configurations and surface effects in confined 
fermionic superfluids as in \cite{ancilotto2013}, where variational results
are well reproduces by density-functional-theory simulations.
By solving the eigenvalue problem of the Hessian matrix 
$\text{Hess}(E) = \frac{\partial^2 E}{ \partial \sigma_i \partial \sigma_j}$
computed in the solitonic (or droplet) minimum, three different
oscillation modes are allowed. In Fig. \ref{fig2} we represent the three
oscillation frequencies ($\omega_I$, $\omega_{II}$, $\omega_{III}$) 
as predicted by the Gaussian ansatz. 

In the solitonic state one frequency, labelled $\omega_I$, can be easily excited,
reflecting the absence of confinement along a direction. The other two modes
are degenerate and converge to $2$ (in units of $\hbar\omega_{\perp}$) as 
in a weakly interacting BEC \cite{salasnich2002}.
The degeneracy of $\omega_{II}$ and $\omega_{III}$ is an obvious consequence of the 
cylindrical symmetry of the radial confinement. 
This structure is clear in the 
panel (d) of Fig. \ref{fig2}, where mode frequencies are 
depicted for the $N=2000$ case: $\omega_{II} = \omega_{III} \simeq 2$, 
two order of magnitude higher than $\omega_I$. The same feature survives 
by increasing the particles number (middle and bottom panel of Fig. \ref{fig2}) in
the solitonic side of the crossover for low values of $\Delta a$.

In the droplet side of the crossover, Fig. \ref{fig2} also shows
the particle-emission-threshold $-\mu$.
The chemical potential $\mu$
is obtained by differentiating the total energy
in Eq. \eqref{energy density with delta} with respect to $N$ within
the Gaussian ansatz in Eq. \eqref{gaussian ansatz}. 
In the solitonic state, $\mu > 0$ for every value of $\Delta a$
and lies above $\omega_I$ and consequently this frequency can be excited 
and experimentally revealed. 
The eigenvector corresponding to $\omega_I$
displays the monopole character of this excitation, since 
$\mathbf{v}_I = \pm \alpha (1,1,1)^T$.
Moving to the droplet-side of the crossover, 
frequencies lie above the emission threshold
in agreement with the prediction of \cite{petrov2015,semeghini2017-arxiv}. 
This implies that excitations of the system
are damped by expelling particles from the droplet in a peculiar 
self-evaporation process.
We also verified that, for $\Delta a=-5.1\,a_0$, as in Fig. \ref{fig2}c, upon increasing the
particle number to larger values one enters into the self-evaporation regime ($5.2 \cdot 10^4<N<1.4\cdot 10^5$). 
Increasing even more the system size, self-evaporation ceases and the excitations
are, at least in principle, observable. It is also interesting to observe that droplet length increases almost linearly with 
the particle number already for $N \ge 4\cdot 10^4$.

\section{Spin-Dipole Oscillations}

In this section we consider the relative dynamics 
between the two components of the mixture. 
We describe the oscillatory motion occurring when the two component
are displaced from the center of the harmonic trap.
To model their collective dynamics, 
we go beyond the assumption where the two components 
occupy the same spatial mode.
We may introduce a weak confinement also along the 
$z$-axis with corresponding aspect ratio $\lambda_z=\omega_z/\omega_\perp$.

Then, we define a center of mass coordinate of each of the two components.
The resulting Gaussian variational ansatz now reads
\begin{equation}
\psi_j = \sqrt{\frac{N_j}{\pi^{3/2}\sigma^2_r \sigma_z}}
e^{-\frac{x^2+y^2}{2\sigma_r^2}} \exp\Big[-\frac{(z-z_j)^2}{2\sigma_z^2}
+i\alpha_j z\Big] \;,
\label{gaussian ansatz 2}
\end{equation}
where the fields $\psi_1$ and $\psi_2$ obey the normalization condition 
$\int d^3\mathbf{r}|\psi_j|^2 = N_j$.
In the equation above we assumed cylindrical 
symmetry $\sigma_x = \sigma_y =\sigma_r$. 
$\lbrace \alpha_i \rbrace_{i= 1,2}$ describe
the corresponding slopes of $\lbrace z_i \rbrace_{i= 1,2}$. 
The set of variational parameters is then given by 
$\lbrace z_1, z_2, \alpha_1, \alpha_2 \rbrace$. 

Replacing \eqref{gaussian ansatz 2} into
the Euclidean action in Eq. \eqref{euclidean action}, we derive the
Lagrangian
\begin{equation}
\begin{aligned}
L &= \sum_j \bigg\lbrace  -\hbar N_j z_j \dot{\alpha_j}
-\frac{\hbar^2N_j}{2m}\alpha_j^2 - \frac{\hbar^2N_j}{2m}
\Big(\frac{1}{\sigma_r^2} +\frac{1}{2\sigma_z^2}\Big) \\
& \qquad -\frac{N_j}{4}m\omega_0^2\big(2\sigma_r^2 +\lambda_z^2 \sigma_z^2 
+2\lambda_z^2 z_j^2\big)-\frac{1}{2}\frac{g_{jj}N_j^2}{2\sqrt{2}
	\pi^{3/2}\sigma_r^2 \sigma_z} \bigg\rbrace \\
& \qquad - \frac{g_{12}N_1 N_2 \exp\Big[-\frac{(z_j-z_k)^2}{2\sigma_z^2}\Big]}
{2\sqrt{2}\pi^{3/2}\sigma_r^2 \sigma_z} + \big(\textup{BMF}\big) \;.
\end{aligned}
\label{lagrangian density}
\end{equation} 
where the quantum fluctuations contribution in Eq. \eqref{lhy correction} 
is simplified by assuming $g_{12}^2 = g_{11}g_{22}$, 
meaning that we are dealing with the system close to
the mean-field instability threshold \cite{stringari-book}. 
The beyond mean field term in the action then reads
\begin{equation}
\big(\textup{BMF}\big)=\Big(\frac{m}{\hbar^2}\Big)^{3/2}
\int d^3\mathbf{r} \big(g_{11}|\psi_1|^2 
+g_{22}|\psi_2|^2\big)^{5/2} \;.
\label{lhy semplificato}
\end{equation}
\begin{figure}[ht!]
\centering
\includegraphics[width=1.0\columnwidth]{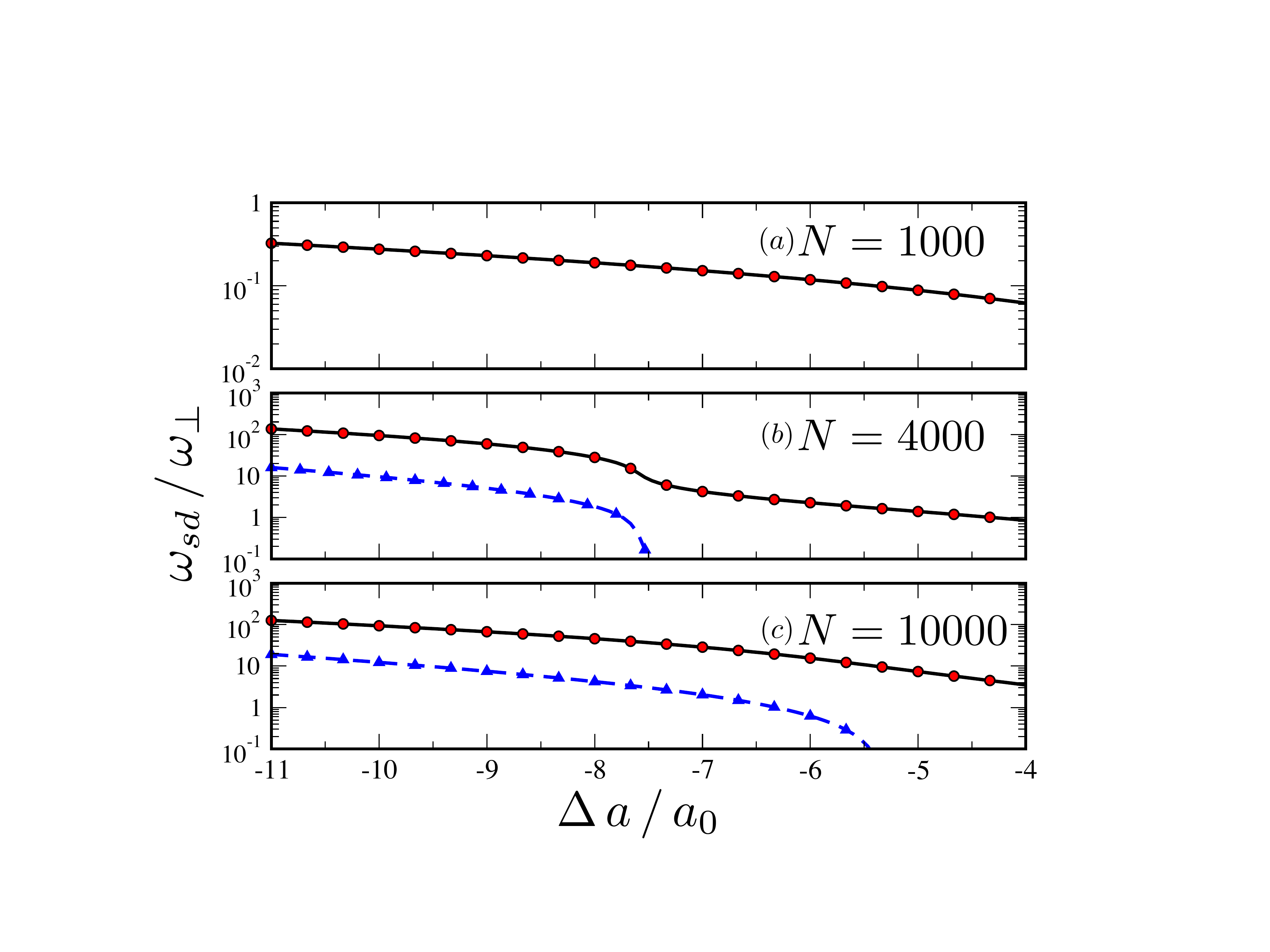}
\caption{\textit{Spin dipole oscillations frequencies}.
(Red dots) Spin dipole oscillations frequency and (Blue triangles) 
single-particle emission threshold $-\mu$ for $N=1000, 4000, 10000$ 
particles. For $N=1000$ the chemical potential is positive in the range
shown in the figure (soliton regime).
In all cases the spin-dipole frequency is higher than $-\mu$ by more
than one order of magnitude.
The plots shown are done with equal intra-species interaction $g_{11}=g_{22}$.
}
\label{fig4}
\end{figure}
In addition, to keep the calculation as analytical as possible, 
we consider a symmetric mixture, 
by taking $g_{11}=g_{22}$ with corresponding equilibrium numbers $N_1=N_2$. 
Given the vector of variational parameters ${\bf q}$,
the Euler-Lagrange equations are computed from
Eq. \eqref{lagrangian density} via
$\frac{\text{d}}{\text{d}t}\frac{\partial L}{\partial \dot{{\bf q}}_i}
- \frac{\partial L}{\partial {\bf q}_i}=0$.
Accordingly, the equation for the slope 
$\alpha_i = m\dot{z}_i/\hbar$
is elementary, so its dynamics it is completely determined
by the one of $z_i$. A closed system of (linearized) diffential equation 
can thus be derived straightforwardly
\begin{equation}
\begin{aligned}
\ddot{z}_1 +\omega_0^2\lambda_z^2 z_1  &= 
\sqrt{\frac{2}{\pi}}\bigg(\frac{a_{12}\hbar^2 N_1}{m^2\sigma_r^2 
	\sigma_z^3}\bigg) (z_1-z_2) \\
&\quad - \frac{1024}{25\pi^{1/4}} \bigg(\frac{a_{11}^{5/2}\hbar^2 N_1^{3/2}}
{m^2 \sigma_r^3 \sigma_z^{7/2}}\bigg)(z_1-z_2) \\
(z_1 \rightleftharpoons z_2) &
\;.\\
\end{aligned} \\
\label{equation of motion for z}
\end{equation}
The equation for $z_2$ has the same structure except for a global sign
on the right side of the equation above. The sum of these two equations
leads to the equation for the longitudinal motion of the center of mass oscillating 
with the trap frequency $\omega_0\lambda_z$ (Kohn's theorem).

Concerning the relative coordinate $\tilde{z} = z_1-z_2$, 
we get
\begin{equation}
\ddot{\tilde{z}}+\omega_{\text{rel}}^2\tilde{z} = 0 \;.
\label{eq. del moto relativo}
\end{equation}
By rescaling the length in units of $a_{\perp}$, the energies
in units of $\hbar\omega_{\perp}$, the frequency $\omega_{rel}$
of the spin-dipole oscillations reads (in unit of $\omega_{\perp}^{-1}$),
\begin{equation}
\begin{aligned}
\frac{{\omega}_{rel}^2}{\omega_{\perp}^2} & = \lambda_z^2 - 
\sqrt{\frac{8}{\pi}}\; \frac{N_1 \big(a_{12}/a_{\perp}\big)}{\sigma_r^2\sigma_z^3}\\
& \qquad + \frac{2048}{25\pi^{1/4}}\; \frac{(a_{11}/a_{\perp})^{5/2} N_1^{3/2}}
{\sigma_r^2 \sigma_z^{7/2}}\;,
\end{aligned}
\label{oscillation frequency rescaled}
\end{equation}
which, in absence of longitudinal confinement 
(i.e. $\lambda_z = 0$) is always a positive defined quantity 
when the effective mean-field interaction is attractive 
($\Delta a \lesssim 0$).
The results of the numerical analysis of the 
spin dipole frequencies are summarized in Fig. \ref{fig4}
for a wide range of interactions and different particle numbers.
We observe that only in the soliton regime do spin oscillations
become observable ($N=1000$). In the droplet phase self-evaporation
is the dominant mechanism.

Concerning Eq. \eqref{oscillation frequency rescaled}, 
we highlight the different sources of the two terms contributing
to spin-dipole oscillations. The first one represents the mean-field attraction
between the two-component and thus depends only on $a_{12}$. Since we are
considering the mutual dynamics between the components, it is reasonable that
purely intra-component terms do not contribute to $\omega_{rel}$. 
One has however to
consider that Gaussian fluctuations are encoded in Eq. \eqref{lhy semplificato} in
such a way that it is not possible to split it in two terms, each one referring
to one component. 

Eq. \eqref{oscillation frequency rescaled} provides another peculiar
feature of self-bound states in binary Bose mixtures: the occurrence
of spin-dipole oscillations is not inhibited by turning off the longitudinal 
confinement, i.e. $\lambda_z=0$. Indeed, the interplay between the mean-field
inter-component attraction and the repulsion arising from quantum fluctuations
establishes a restoring force.
Then, differently from the
mean-field scenario depicted in \cite{ferrari2016}, the presence of a longitudinal
confinement is not essential for
spin-dipole oscillations to take place due to the attractive nature of the
mean-field interaction between the two species. 

\section{Experimental numbers}
By now, experiments on self-bound states of binary Bose mixtures
were performed by considering 
a mixture of $^{39}$K atoms in two different internal states,
$\ket{1}=\ket{F=1, m_F = -1}$ and $\ket{2}=\ket{F=1, m_F = 0}$. 
By tuning an external magnetic field, Feshbach resonances provide
the possibility of exploring a wide range of intra and inter-particles
scattering lengths. More precisely \cite{derrico2007}, by
tuning the magnetic field between $B = 54 \div 57.5$ G, 
where $a_{11} = 33.5\, a_0$ and $a_{12} = -53.6\, a_0$, 
the variation of $a_{22}$ provides a whole range of values for
$\Delta a$, from $-15\, a_0$ to $+10\, a_0$.
Collective oscillations in binary mixtures can be probed
in an experimental setup similar to the
one developed in \cite{tarruell2017-arxiv}. The existence of the soliton-droplet
crossover can be revealed by engineering a cigar-shaped mixture and
then removing the trapping along the $z$-axis and observing that the system
still has a well-defined longitudinal length. 

Our analytical results are compatible with the occurrence of self-evaporation
for a large interval of experimental parameters. 
Only in a small window in the parameters space
do we have that where monopole/quadrupole collective oscillations can be
in principle observed as highlighted in Fig. \ref{fig3}. 
Moreover, we developed our analysis in a
zero-temperature framework, leaving unanswered the question about what happens
for small, but finite temperatures.
 
In order to test our analytical results concerning the spin dipole oscillations
in Eq. \eqref{oscillation frequency rescaled} and Fig. \ref{fig3},
we require a tighter range of experimental values. In particular,
as in \cite{cappellaro2017}, we consider the symmetric mean-field ground state
given by the condition $a_{11} \simeq a_{22} \simeq 33.5 \, a_0$ at 
$B = 54.5 \, G$, where $a_{12} = -54 \, a_0$. 
After a careful tuning of intra- and inter-component scattering lengths, 
the experimental protocol to probe spin-dipole oscillations is reported in
\cite{ferrari2016}. Indeed, each components in an overlapping initial state 
can be displaced by applying a magnetic field gradient $\delta B$ along
the longitudinal axis, generated by two coils in anti-Helmoltz configuration.
Also, as stated in the previous section, a longitudinal
trapping is not mandatory to experimentally probe spin-dipole oscillations,
because of the peculiar structure of Gaussian fluctuations contribution to
the energy density and the attractive mean-field interaction.
Then we emphasize that from the experimental perspective
our study may provide insights for the observation of spin dipole oscillations
of solitons in a parameter regime which is complementary to that of 
\cite{ferrari2016}.
Finally, we mention that the effect of three-body losses could be included
into a full description of the excitation dynamics. This can done 
with the inclusion of a term $-i\hbar \frac{L_3}{2}|\psi|^4$ in the right side of the 
Gross-Pitaevskii equation \cite{Wenzel17}. A dedicated study will be performed in a future
publication.

\section{Conclusions}
In this work we analyzed the behavior of collective modes of two-component BECs
with attractive interparticle interactions along the 
crossover from soliton to self-bound droplets
in a quasi one dimensional waveguide.
We observed a sharp difference in the collective modes in the two regimes. 
In the soliton regime
modes vary smoothly upon the variation of particle number or interaction strength.
On the droplet side collective modes are inhibited by the
 emission of particles. This mechanism turns out to 
be dominant for a wide range of particle numbers and interactions. 
In a small window of particle number
range and for intermediate interactions we find that monopole 
frequency is likely to be observed.
In the last part we have studied the spin-dipole modes 
for the case of equal intraspecies interactions and
equal equilibrium particle numbers in the presence 
of a weak longitudinal confinement.
We have found that such modes might be again unobservable 
as their frequency is higher than 
the particle emission spectrum by at least one order of magnitude in the droplet phase.
In conclusion, we have found that collective modes provide a 
complementary way to characterize two-component
BECs in an optical waveguide. 
Moreover they provide a strong signature to distinguish between soliton-like
configurations from self-bound droplets.
We expect that our work, inspired by experiments in the Barcelona group, 
may provide a useful indication
for further experiments with multi-component BECs.

\section*{Acknowledgements}

T. M. acknowledges CNPq for support through Bolsa de produtividade 
em Pesquisa n. 311079/2015-6 and the hospitality of
Physics Department of the University of Padova and the University of L'Aquila where 
part of the work was done.
L.S. thanks for partial support the 2016 BIRD project 
\textit{Superfluid properties of Fermi gases in optical potentials}
of the University of Padova.

\end{document}